\documentclass[conference,10pt]{IEEEtran}

\usepackage[cmex10]{amsmath}
\usepackage{amssymb}
\usepackage{amsfonts}
\usepackage{amsthm}
\usepackage{array}
\usepackage{dsfont}
\usepackage{color}
\usepackage{graphicx}
\usepackage{epstopdf}
\usepackage{cite}
\usepackage[caption=false]{subfig}

\long\def\symbolfootnote[#1]#2{\begingroup%
\def\thefootnote{\fnsymbol{footnote}}\footnote[#1]{#2}\endgroup}
\allowdisplaybreaks[1]

\IEEEoverridecommandlockouts

\begin{document}

\title{Optimal Demand-Side Management for Joint Privacy-Cost Optimization with Energy Storage}

\author {
  \IEEEauthorblockN{Giulio Giaconi and Deniz G\"{u}nd\"{u}z}

  \IEEEauthorblockA{Imperial College London,  London, UK\\
    {\{g.giaconi, d.gunduz\}}{@imperial.ac.uk}
\and
\IEEEauthorblockN{H. Vincent Poor}
\IEEEauthorblockA{Princeton University, Princeton, NJ, USA \\
poor@princeton.edu}
}
\thanks{The first author gratefully acknowledges the Engineering and Physical Sciences Research Council (EPSRC) of the UK for funding his PhD studies (award reference \#1507704). This work was also supported in part by the EPSRC through the project COPES (\#173605884), by the Miller Institute for Basic Research in Science, University of California, Berkeley, and by the U.S. National Science Foundation under Grant ECCS-1549881.}
}

\maketitle

\begin{abstract}
The smart meter (SM) privacy problem is addressed together with the cost of energy for the user. It is assumed that a storage device, e.g., an electrical battery, is available to the user, which can be utilized both to achieve privacy and to reduce the energy cost by modifying the energy consumption profile. Privacy is measured via the mean squared-error between the SM readings, which are reported to the utility provider (UP), and a target load; while time-of-use pricing is considered for energy cost calculation. The optimal trade-off between the achievable privacy and the energy cost is characterized by taking into account the limited capacity of the battery as well as the capability to sell energy to the UP. Extensive numerical simulations are presented to evaluate the performance of the proposed strategy for different system settings.
\end{abstract}
\IEEEpeerreviewmaketitle

\section{Introduction}

\emph{Smart meters} (SMs) are the key component of the next generation of power grids, the so-called \emph{smart grid}, since they enable the monitoring of a user's power consumption in almost real time. SMs allow a utility provider (UP), i.e., the entity that sells energy to customers, to determine the electricity cost dynamically, and a distribution system operator (DSO), i.e., the entity that manages the power grid, to better manage the grid itself. In addition, SMs allow users to better monitor their electricity consumption, integrate renewable energy sources into the grid, and sell surplus energy to the UP.

However, because of the high accuracy and granularity of the SM readings, SMs also entail serious privacy implications. By using non-intrusive appliance load monitoring techniques it is possible to infer user's activities and behaviors, and determine, for example, a user's presence at home, her religious beliefs, disabilities, illnesses, and even which TV channel she is watching \cite{Rouf:2012}. To address these issues, several methods have been proposed so far, including data obfuscation \cite{Bohli:2010}, data aggregation \cite{Li:2011}, and data anonymization \cite{Petrlic:2010}. However, a disadvantage of these approaches is that the DSO could potentially embed additional sensors to monitor the energy requested by a user, without fully relying on SM readings. Additionally, data obfuscation methods misreport the real energy consumption, thus preventing the DSO from accurately monitoring the grid states and rapidly reacting to problems such as outages or energy theft. Another approach is to directly modify the user's energy consumption, called the \emph{input load}, rather than simply modifying the data reported to the UP, called the \emph{output load}. Thus, the aim is to make the output load as different from the input load as possible, by, for example, filtering energy via a rechargeable energy storage device, i.e., a battery, as studied in \cite{Kalogridis:2011,Yang:2012,Tan:2013,Li:2016,Giaconi:2017arxivTIFS,Tan:2017}.

In this paper, we adopt the latter approach and consider the presence of a battery which can store energy from the grid to partially satisfy a user's energy demand. Our goal is to minimize the energy cost for the user while keeping the information leakage to the energy provider minimal. While a widely accepted definition of privacy is still elusive, one can suggest that privacy is achieved in a smart metering system when it is not possible to distinguish a specific appliance load from the total power consumption \cite{Kalogridis:2010}. From this perspective, a high degree of privacy can be achieved by flattening the power consumption around a constant value, as considered in \cite{Tan:2017}. However, a completely constant consumption may not be practically viable and convenient, since, for example, the cost may vary greatly during the system operation due to time-of-use tariffs. Also, it is not clear what the advantage of such a policy would be, as compared, for example, to the case of having multiple levels of constant consumption.

For these reasons, in this paper we consider a more general and flexible target load profile compared to \cite{Tan:2017}, and jointly minimize the cost of energy and the mean squared-error between the SM readings and the  target load. Moreover, we also consider the possibility of selling energy to the UP as a way to further lower the cost and improve the privacy. We impose the target load to remain constant over each price period, since it is reasonable to assume that a user may prefer to draw more energy during off-peak periods (when the price is lower) as compared to the peak price period, and this behavior would not necessarily leak more information about the user.

The remainder of the paper is organized as follows. In Section \ref{sec:SystemModel} we present the system model, while in Section \ref{sec:targetPiecewiseConstant} we analyze the proposed target load analytically. Numerical results are presented in Section \ref{sec:numResults}, while conclusions are drawn in Section \ref{sec:conclusion}.

\begin{figure}[!t]
\centering
\includegraphics[width=1\columnwidth]{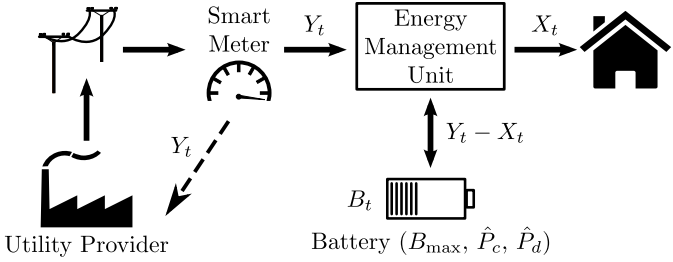}
\caption{The system model. $X_t$, $Y_t$ and $Y_t-X_t$ are the user input load, the output load, and the energy exchanged with the battery at time $t$, respectively. The dashed line represents the meter
readings being reported to the UP.}
\label{fig:SystemModel}
\end{figure}

\section{System Model}\label{sec:SystemModel}



We consider the discrete time system model depicted in Fig. \ref{fig:SystemModel}, where each time index $t$ represents a time slot (TS) for $1 \leq t \leq N$. For each TS $t$, the input load is $X_t \in \mathcal{X}$ and the output load is $Y_t \in \mathcal{Y}$. We assume the presence of a rechargeable energy storage device, i.e., a battery, of capacity $B_{\max}$, which is empty at $t=0$ and can be used both to filter the input load to provide privacy to the user, and to shift energy intake from the grid to minimize cost. The energy management unit determines how much energy to draw from the grid, $Y_t$, and how much energy to exchange with the battery, $Y_t-X_t$. The time duration between two consecutive measurements is $\delta_i \triangleq t_i-t_{i-1}$, which is assumed to be constant for the sake of simplicity. The total working time of the system is thus $T=\delta N$, and $X_{t}$ is the user's energy demand occurring in time interval $[(t-1)\delta, t\delta]$. Let $B_t\in [0, B_{\max}]$ denote the battery state-of-charge (SOC) at the end of TS $t$. The battery is charging if $Y_t-X_t>0$, discharging if $Y_t-X_t<0$, or remains in the same SOC if $Y_t-X_t=0$. We do not allow wasting grid energy; that is, there are no battery overflows, i.e.,
\begin{equation} \label{eq:batteryConstraint}
\sum_{\tau=1}^{t} (Y_{\tau}-X_{\tau}) \delta \leq B_{\max}, \qquad t=1,2,\ldots N.
\end{equation}

While additional energy can be stored in the battery for future use, we do not allow energy rescheduling or demand shifting, i.e., user's energy requests are always satisfied as they occur:
\begin{equation} \label{eq:energySatisfied}
\sum_{\tau=1}^{t} Y_{\tau} \geq \sum_{\tau=1}^{t} X_{\tau}, \qquad t=1,2,\ldots N.
\end{equation}

In particular, (\ref{eq:energySatisfied}) is also implicitly verified by the battery SOC evolution equation:
\begin{equation}
B_{t+1} = B_{t} + Y_{t+1} - X_{t+1}.
\end{equation}

A further constraint is represented by the maximum amount of power with which the battery can be charged or discharged, denoted by $\hat{P}_{c}$ and $\hat{P}_{d}$, respectively. Thus, we have
\begin{align}
Y_t &\leq X_t + \hat{P}_{c}, \qquad t=1,2,\ldots N, \label{eq:peakPowerCharging}\\
Y_t &\geq X_t - \hat{P}_{d}, \qquad t=1,2,\ldots N. \label{eq:peakPowerDischarging}
\end{align}

Here we assume that the battery has perfect efficiency, i.e., no leakages, to understand the optimal cost-privacy trade-off in an ideal setting. The effects of battery inefficiencies and degradations may be taken into account by introducing a constant rate leakage for the energy stored in the battery, or by imposing a further cost whenever the battery is used, within the framework studied here.

At first we do not consider selling energy to the UP, i.e., the energy drawn from the grid is always non-negative
\begin{equation} \label{eq:ouputNonnegative}
Y_t \geq 0, \qquad t=1,2,\ldots N.
\end{equation}

Later we will lift this constraint. Given $(X_t, B_t)=(x_t, b_t)$, $B_{\max}$ and the constraints (\ref{eq:peakPowerCharging})-(\ref{eq:ouputNonnegative}), the set of feasible energy requests at time $t$ is
\begin{multline}\label{eq:feasibleSetY}
\bar{\mathcal{Y}_t}(x_t,b_t) \triangleq \Big\{ y_t \in \mathcal{Y}: \\
\Big[x_t-\min\{b_t,\hat{P}_d\}\Big]^+ \leq y_t \leq x_t + \min\{\hat{P}_c,B_{\max}-b_t\}\Big\},
\end{multline}
where $[a]^+=a$ if $a>0$, and $0$ otherwise.

When $t=N$ no energy should be left unused in the battery; that is, we impose
\begin{equation} \label{eq:totalEnergy}
\sum_{t=1}^{N} Y_t = \sum_{t=1}^{N} X_t.
\end{equation}

We aim at designing an \emph{energy management policy} that decides on the amount of energy to request from the UP at each TS $t$ while satisfying (\ref{eq:energySatisfied})-(\ref{eq:totalEnergy}). We consider an offline setting, in which the input load $X^N$ and the cost of energy $C^N$ are known beforehand and our goal is to jointly minimize the information leaked about the user's energy consumption as well as the cost the user incurs to purchase energy from the UP. While non-causal knowledge of the price is a realistic assumption in today's energy networks, non-causal knowledge of power consumption is valid for appliances such as refrigerators, boilers, heating and electric vehicles, whose energy consumption can be accurately predicted over certain finite time frames. Energy management with imperfect input load predictions will be studied in a future work. We also remark that the solution of the offline problem yields meaningful lower bounds on the performance of the more general online setting. In fact, in the offline setting, the knowledge of future energy consumption enables the user to better match the output with the target load, thus leading to the minimum amount of information leaked to the UP.

We measure privacy as the mean squared-error between $Y^N$ and a target load $W^N$ over $T$ TSs, i.e.,
\begin{equation}\label{eq:privacy}
\mathcal{P} \triangleq \frac{1}{N}\sum_{t=1}^N (Y_t - W_t)^2,
\end{equation}
according to which, perfect privacy is achieved when $Y_t = W_t$, $\forall t$. When energy cannot be sold to the UP, $W_t$ is non-negative, i.e.,
\begin{equation} \label{eq:targetNonnegative}
W_t \geq 0, \qquad t=1,2,\ldots N.
\end{equation}

The average cost incurred by the user during time $T$ is
\begin{equation}\label{eq:cost}
\mathcal{C} \triangleq \frac{1}{N} \sum_{t=1}^N C_t  Y_t ,
\end{equation}
\begin{figure*}[t]
\begin{multline}\label{eq:lagrangian}
  \mathcal{L}(x,\lambda,\nu) = \frac{1}{N} \sum_{i=1}^{M} \Bigg[ \sum_{t=t_{c^{(i-1)}}}^{t_{c^{(i)}}} \alpha \big(Y_t - W^{(i)}\big)^2 + (1-\alpha) Y_t C^{(i)} \Bigg] +  \sum_{t=1}^{N} \lambda_{t}^{(1)} \sum_{\tau=1}^{t} (X_{\tau}-Y_{\tau}) +  \sum_{t=1}^{N} \lambda_{t}^{(2)} \bigg\{ \bigg[\sum_{\tau=1}^{t} (Y_{\tau}-X_{\tau})\delta \bigg] - B_{\max}\bigg\} \\
  +  \sum_{t=1}^{N} \lambda_{t}^{(3)} (Y_t - X_t - \hat{P}_c) + \sum_{t=1}^{N} \lambda_{t}^{(4)} (X_t -Y_t - \hat{P}_d) -  \sum_{t=1}^{N} \lambda_{t}^{(5)} Y_t -  \sum_{i=1}^{C_M} \lambda_{i}^{(6)} W^{(i)} + \nu^{(1)} \sum_{t=1}^{N}  (Y_t-X_t).
\end{multline}
\hrulefill
\vspace{-3mm}
\end{figure*}
where $C_t$ is the cost of energy at time $t$, which is determined by the specific time-of-use tariff employed by the UP.

Compared to \cite{Tan:2017} where the target load is set to be a constant function over time, in the following section we consider a more general target load, specifically a function which assumes a constant value for each price period.

\section{Target Load as a Piecewise Constant Function}\label{sec:targetPiecewiseConstant}

In this section the objective of the optimization problem is to determine both the optimal target $\{W^*\}_{t=1}^N$ and the optimal output load $\{Y^*\}_{t=1}^N$, so that the average squared-distance between them is minimized. The target load is set to be piecewise constant over price periods. We argue that such an output load trades off privacy with cost. Multiple target values for energy intake from the grid intuitively reveal more information about the user's real energy consumption behavior compared to a single constant target value. However, it also provides more flexibility to reduce the cost and to match the target load, providing a trade-off between cost and privacy.

\begin{figure}[!t]
\centering
\includegraphics[width=0.9\columnwidth]{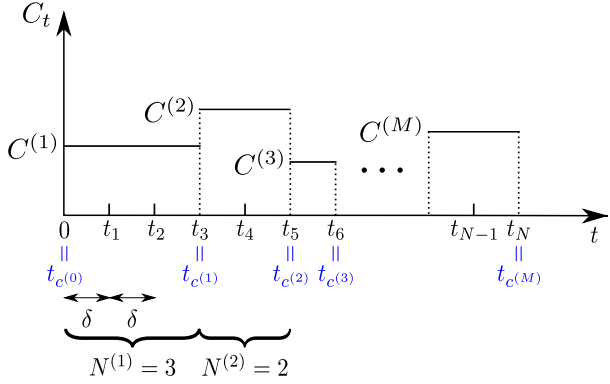}
\caption{Time-of-use tariff. $t_{c^{(i)}}$, for $i=1,\ldots,M-1$, are the time instants at which the price of energy changes, while $t_{c^{(0)}}=0$ and $t_{c^{(M)}}=t_N$ are the extremes of the time interval considered.}
\label{fig:energyCost}
\end{figure}

Let $C^{(i)}$ be the cost of the energy purchased from the UP during the $i$-th price period, where $1 \leq i \leq M$ and $M$ is the total number of price periods during time $T$. The $i$-th price period spans from $t_{c^{(i-1)}}$ to $t_{c^{(i)}}$, and $N^{(i)} \triangleq (t_{c^{(i)}}-t_{c^{(i-1)}})/\delta$ is the number of TSs contained within price period $i$. Fig. \ref{fig:energyCost} depicts this model. Let $W^{(i)}$ be the constant target load for the $i$-th time period. Given the nature of the objective functions and the constraints, pairs of ($\mathcal{P}$, $\mathcal{C}$) form a convex region and the optimal points can be characterized by the Pareto boundary of this region. Then, our problem can be cast as a weighted combination of privacy (\ref{eq:privacy}) and cost (\ref{eq:cost}), i.e.,
\begin{equation}\label{eq:convexOptConstantPiecewise}
\min_{Y_t,W^{(i)}}  \frac{1}{N} \sum_{i=1}^{M} \Bigg[ \sum_{t=t_{c^{(i-1)}}}^{t_{c^{(i)}}} \alpha (Y_t - W^{(i)})^2 + (1-\alpha) Y_t C^{(i)} \Bigg],
\end{equation}
where $0 \leq \alpha \leq 1$ is the weighting parameter between privacy and cost, according to which if $\alpha=0$ only cost of energy is minimized, whereas if $\alpha=1$ only information leakage is minimized.

Based on (\ref{eq:convexOptConstantPiecewise}) and the constraints (\ref{eq:batteryConstraint})-(\ref{eq:energySatisfied}), (\ref{eq:peakPowerCharging})-(\ref{eq:ouputNonnegative}), (\ref{eq:totalEnergy}) and  (\ref{eq:targetNonnegative}), we define the Lagrangian function as (\ref{eq:lagrangian}) shown at the top of the page, where $\lambda_{t}^{(j)} \geq 0$ for $1\leq j\leq 5$ and $1\leq t \leq N$, $\lambda_{i}^{(6)} \geq 0$ for $1\leq i\leq C_M$, and $\nu^{(1)} \geq 0$ are the Lagrange multipliers. The corresponding slackness conditions are imposed on the inequality constraints:
\begin{align}
  &\lambda_{t}^{(1)} \sum_{\tau=1}^{t} (X_{\tau}-Y_{\tau}) = 0, \qquad t = 1, \ldots, N, \\
  &\lambda_{t}^{(2)} \bigg\{ \bigg[ \sum_{\tau=1}^{t} (Y_{\tau}-X_{\tau})\delta \bigg] - B_{\max}\bigg\}=0, \quad t=1\ldots,N, \\
  &\lambda_{t}^{(3)} (Y_t - X_t - \hat{P}_c)=0, \qquad t = 1, \ldots, N, \\
  &\lambda_{t}^{(4)} (X_t -Y_t - \hat{P}_d)=0, \qquad t = 1, \ldots, N, \\
  &\lambda_{t}^{(5)} Y_t=0, \qquad t = 1, \ldots, N, \\
  &\lambda_{i}^{(6)} W^{(i)}=0, \qquad i = 1, \ldots, C_M.   \label{eq:targetLagrangian}
\end{align}

We then apply the Karush Kuhn Tucker (KKT) conditions and set the gradient of the Lagrangian to zero, i.e.,
\begin{align}
\frac{\partial \mathcal{L}}{\partial Y_t} &= \frac{2 \alpha (Y_t-W^{(i)})}{N}   + \frac{(1-\alpha)C^{(i)}}{N} + \sum_{\tau=t}^{N}(\lambda_{\tau}^{(2)}\delta-\lambda_{\tau}^{(1)}) \nonumber \\
&\qquad \qquad + \lambda_t^{(3)}-\lambda_t^{(4)} - \lambda_t^{(5)}+\nu^{(1)} = 0, \label{eq:KKT1}\\
\frac{\partial \mathcal{L}}{\partial W^{(i)}} &= -\sum_{t=t_{c^{(i-1)}}}^{t_{c^{(i)}}} \frac{2 \alpha (Y_t-W^{(i)})}{N} -\lambda_i^{(6)}  = 0 \label{eq:KKT2}.
\end{align}

From (\ref{eq:KKT1}), we obtain the optimal value for $Y_t$ as
\begin{equation}\label{eq:optimalY}
Y^{*}_t= \Bigg[ \frac{N a_t}{2\alpha} + W^{(i)*} - \tilde{C}^{(i)} \Bigg]^{+}, \quad t = 1, \ldots, N,
\end{equation}
where $a_t=\sum_{\tau=t}^{N}(\lambda_{\tau}^{(1)}-\lambda_{\tau}^{(2)}\delta) - \lambda_{t}^{(3)} + \lambda_{t}^{(4)} + \lambda_{t}^{(5)} - \nu^{(1)}$ and $\tilde{C}^{(i)}=\frac{(1-\alpha)C^{(i)}}{2\alpha}$. From (\ref{eq:KKT2}), we obtain the optimal value for $W^{(i)}$ as
\begin{equation}\label{eq:optimalWi}
W^{(i)*}= \Bigg[ \frac{N \lambda_i^{(6)}}{2 \alpha N^{(i)}}  + \frac{\sum_{t=t_{c^{(i-1)}}}^{t_{c^{(i)}}} Y^*_t}{N^{(i)}}  \Bigg]^+, \quad 1 \leq i \leq M.
\end{equation}

The optimal solution (\ref{eq:optimalY}) resembles the classical water-filling algorithm that determines the optimal power allocation for Gaussian parallel channels under a total power constraint \cite{Cover:1991}. However, differently from the classical water-filling formulation, here the water level, $Y^{*}_t + \tilde{C}^{(i)} = \frac{N a_t}{2\alpha} + W^{(i)*} $, is not constant, but varies over time due to the instantaneous constraints. The optimal target for price period $i$, (\ref{eq:optimalWi}), is a function of the Lagrange multipliers and of the mean of the optimal output loads in the same price period.

\begin{figure}
\centering
\subfloat[][$B_{\max,1}=4$.]
{\includegraphics[width=.48\columnwidth]{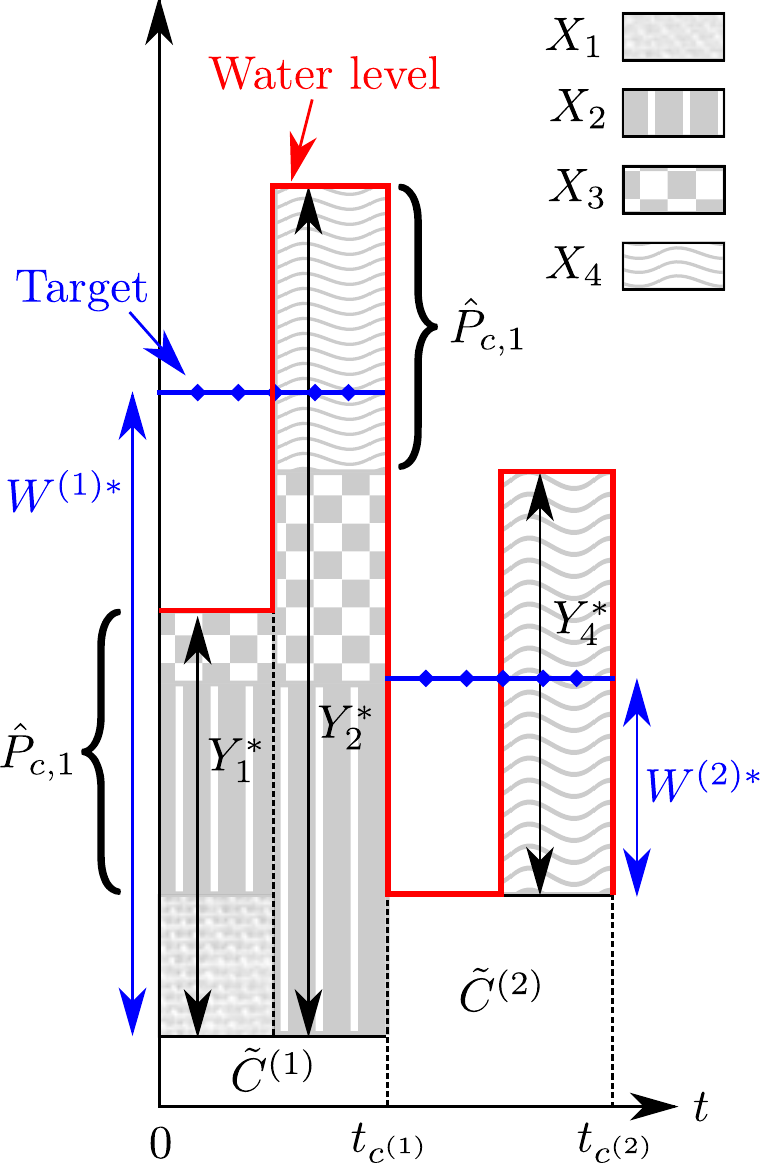}\label{fig:piecewiseEx1}}
\hspace{0.1cm}
\subfloat[][$B_{\max,2}=8$.]
{\includegraphics[width=.48\columnwidth]{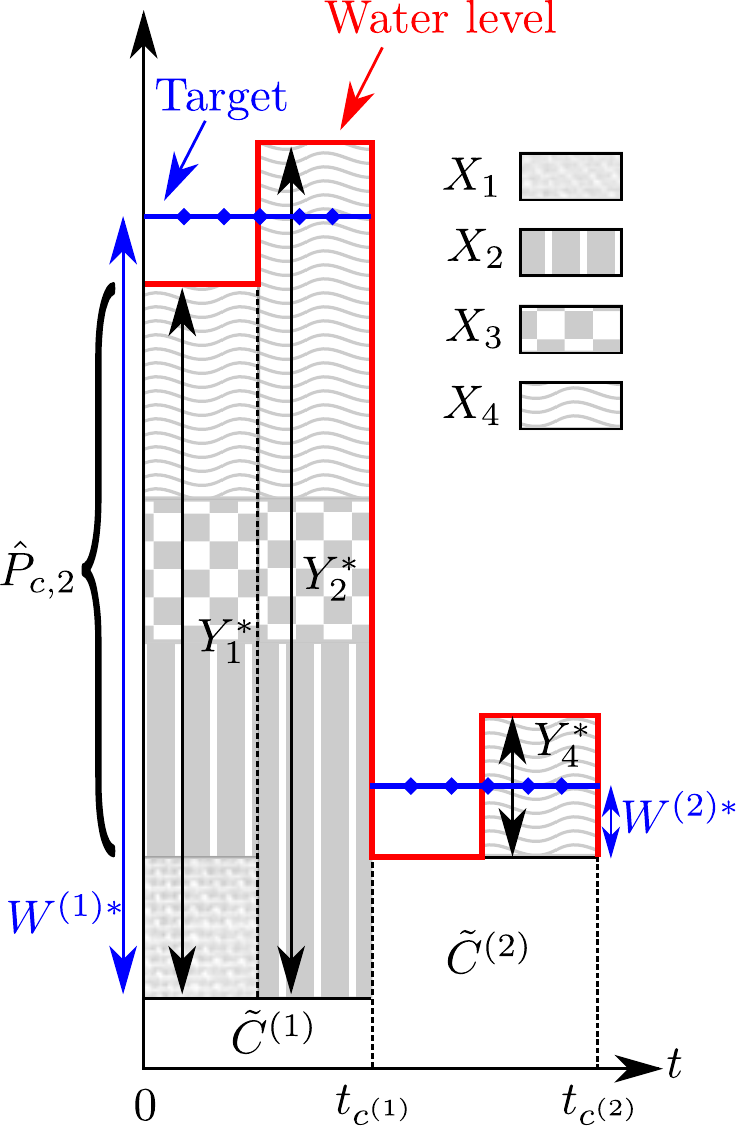}\label{fig:piecewiseEx2}}
\caption{A simplified scenario with a piecewise constant target load.}
\label{fig:piecewiseEx}
\end{figure}

The simple scenario illustrated in Fig. \ref{fig:piecewiseEx} offers some intuition about the optimal solution and its water-filling interpretation. We consider two price periods, $C^{(1)}=1$ and $C^{(2)}=3$, two batteries of parameters $B_{\max,1}=4$ kWh, $\hat{P}_{c,1}=\hat{P}_{d,1}=2$ kW and $B_{\max,2}=8$ kWh, $\hat{P}_{c,2}=\hat{P}_{d,2}=4$ kW, respectively, $N=4$ and $\alpha=0.5$. The input load is $X^4=[1,4,2,5]$ kW. For the sake of simplicity, we consider $\delta_i=1$, so that power and energy can be used interchangeably. Fig. \ref{fig:piecewiseEx1}, shows the optimal solution and water levels for $B_{\max,1}$. Since $C^{(1)}<C^{(2)}$, more energy is requested from the grid during the first price period and stored in the battery to partially satisfy the user's energy consumption during the second price period. However, the energy that can be stored is limited not only by the battery capacity $B_{\max}$ but also by the peak power constraint $\hat{P}_{c,1}$ (\ref{eq:peakPowerCharging}). In the first TS it is indeed $\hat{P}_{c,1}$ that limits the output load. We have $Y^*_1=X_1+\hat{P}_{c,1}$ and the battery SOC at the end of the TS is $B_1=\hat{P}_{c,1}=2$. The water level for $t=1$ is given by $Y_1^* + \tilde{C}^{(1)} = \frac{4 a_1}{2\alpha} + W^{(1)*}$, where $a_1=\sum_{\tau=1}^{4}(\lambda_{\tau}^{(1)}-\lambda_{\tau}^{(2)}\delta) - \lambda_{1}^{(3)} + \lambda_{1}^{(4)} + \lambda_{1}^{(5)} - \nu^{(1)}$, and $W^{(1)*}$ is computed jointly with $Y_1^{*}$ and $Y_2^{*}$ (\ref{eq:optimalWi}). In the second TS, the output load is limited by both the charging peak power constraint and the battery capacity, i.e., $Y^*_2=X_2+\hat{P}_{c,1}$, and $B_2=B_{\max}=4$. In the third TS, $Y^*_3=0$, which does not violate the discharging peak power constraint as $X_3=\hat{P}_{d,1}=2$, and $B_3=2$. In the last TS, $Y^*_4=X_4-\hat{P}_{d,1}=3$ and the battery becomes empty. Fig. \ref{fig:piecewiseEx2} shows that by increasing $B_{\max}$, $\hat{P}_c$ and $\hat{P}_d$, it is possible to further reduce the variance of the output load from the target and the cost of energy.


\subsection{Selling Energy to the Grid}\label{sec:energySoldUPpiecewise}

Here we consider the possibility of selling energy to the UP in order to further improve the user's privacy and save costs. By allowing this, the UP can practically utilize user devices as distributed energy storage. Thus, we lift constraints (\ref{eq:ouputNonnegative}) and (\ref{eq:targetNonnegative}), and we have
\begin{equation}
Y_t :
\begin{cases}
  > 0 , & \mbox{if } \text{energy is purchased from grid}, \\
  < 0 , & \mbox{if } \text{energy is sold to grid}.
\end{cases}
\end{equation}

Without loss of generality, the price of selling energy to the grid is equal to the price of buying energy from it, i.e., we implement the \emph{net metering} approach, according to which the user only needs a single SM, which can measure bi-directional energy flows \cite{Payne:2000}.

The set of feasible energy requests at time $t$ is given by
\begin{multline}\label{eq:feasibleSetYnegative}
\tilde{\mathcal{Y}_t}(x_t,b_t) \triangleq \Big\{ y_t \in \mathcal{Y}: \\
x_t-\min\{b_t,\hat{P}_d\} \leq y_t \leq x_t + \min\{\hat{P}_c,B_{\max}-b_t\}\Big\}.
\end{multline}

The optimization problem is expressed as (\ref{eq:convexOptConstantPiecewise}) and the Lagrangian is similar to (\ref{eq:lagrangian}) without considering the constraints corresponding to $\lambda^{(5)}_t$ and $\lambda^{(6)}_i$. The slackness conditions follow similarly, and by applying the KKT conditions, we obtain the following optimal values for $Y_t$ and $W^{(i)}$:
\begin{align}\label{eq:optimalY_sold}
Y^{*}_t &= \frac{N a_t}{2\alpha} + W^{(i)*} - \tilde{C}^{(i)}, \qquad \forall t, \\
W^{(i)*} &= \frac{\sum_{t=t_{c^{(i-1)}}}^{t_{c^{(i)}}} Y^*_t}{N^{(i)}}, \quad \forall i, \label{eq:optimalWi_sold}
\end{align}
where $a_t=\sum_{\tau=t}^{N}(\lambda_{\tau}^{(1)}-\lambda_{\tau}^{(2)}\delta) - \lambda_{t}^{(3)} + \lambda_{t}^{(4)} - \nu^{(1)}$.

\begin{figure*}[!htbp]
\centering
\subfloat[][Constant target, $\alpha=0.1$.]
{\includegraphics[width=.4\textwidth]{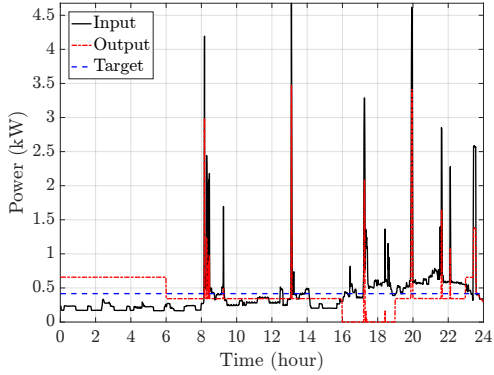}\label{fig:profilesConstant01_B4}}
\subfloat[][Piecewise constant target, $\alpha=0.1$.]
{\includegraphics[width=.4\textwidth]{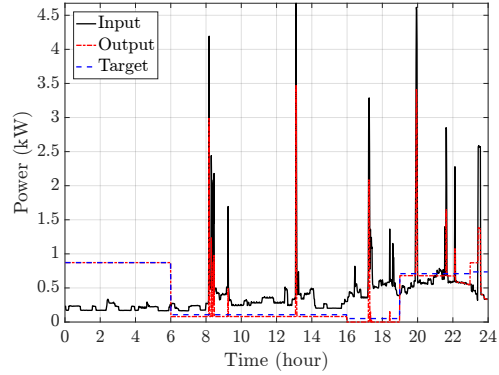}\label{fig:profilesPiece01_B4}}\\
\vspace{-0.3cm}
\centering
\subfloat[][Constant target, $\alpha=0.9$.]
{\includegraphics[width=.4\textwidth]{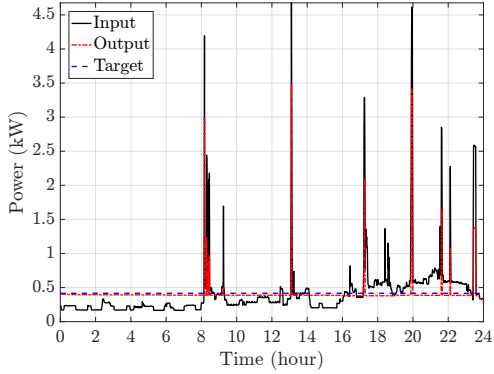}\label{fig:profilesConstant09_B4}}
\subfloat[][Piecewise constant target, $\alpha=0.9$.]
{\includegraphics[width=.4\textwidth]{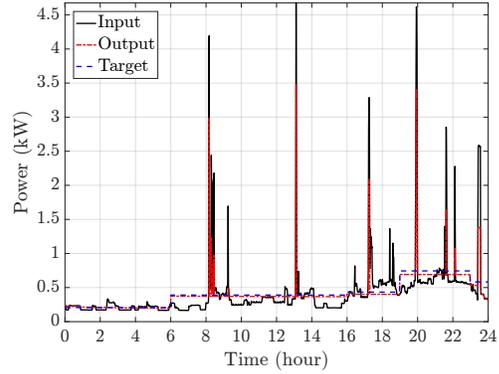}\label{fig:profilesPiece09_B4}}
\caption{Input, target and output loads for a Powervault G200-LI-4KWH battery \cite{powervault}, $\alpha=0.1$ and $\alpha=0.9$.}
\label{fig:profiles}
\vspace{-0.7cm}
\end{figure*}

\section{Numerical Results} \label{sec:numResults}

\begin{table}
\caption{Specifications of the batteries considered in the simulations.}
\centering
\resizebox{\columnwidth}{!}{
\begin{tabular}{ |c|c|c|c| }
\hline
Residential Battery & \begin{tabular}[x]{@{}c@{}} Capacity \\ (kWh) \end{tabular} & \begin{tabular}[x]{@{}c@{}}Charging Peak \\ Power (kW) \end{tabular} & \begin{tabular}[x]{@{}c@{}} Discharging Peak \\ Power (kW) \end{tabular}\\
\hline
Powervault G200-LI-4KWH \cite{powervault}& $4$ & $1.2$ & $1.4$\\
\hline
Tesla Powerwall $2$ \cite{tesla}& $13.5$ & $5$ & $5$\\
\hline
\end{tabular}}
\label{tab:batteryCapacity}
\end{table}

\begin{figure}
\centering
\subfloat[][Energy cannot be sold.]
{\includegraphics[width=.5\columnwidth]{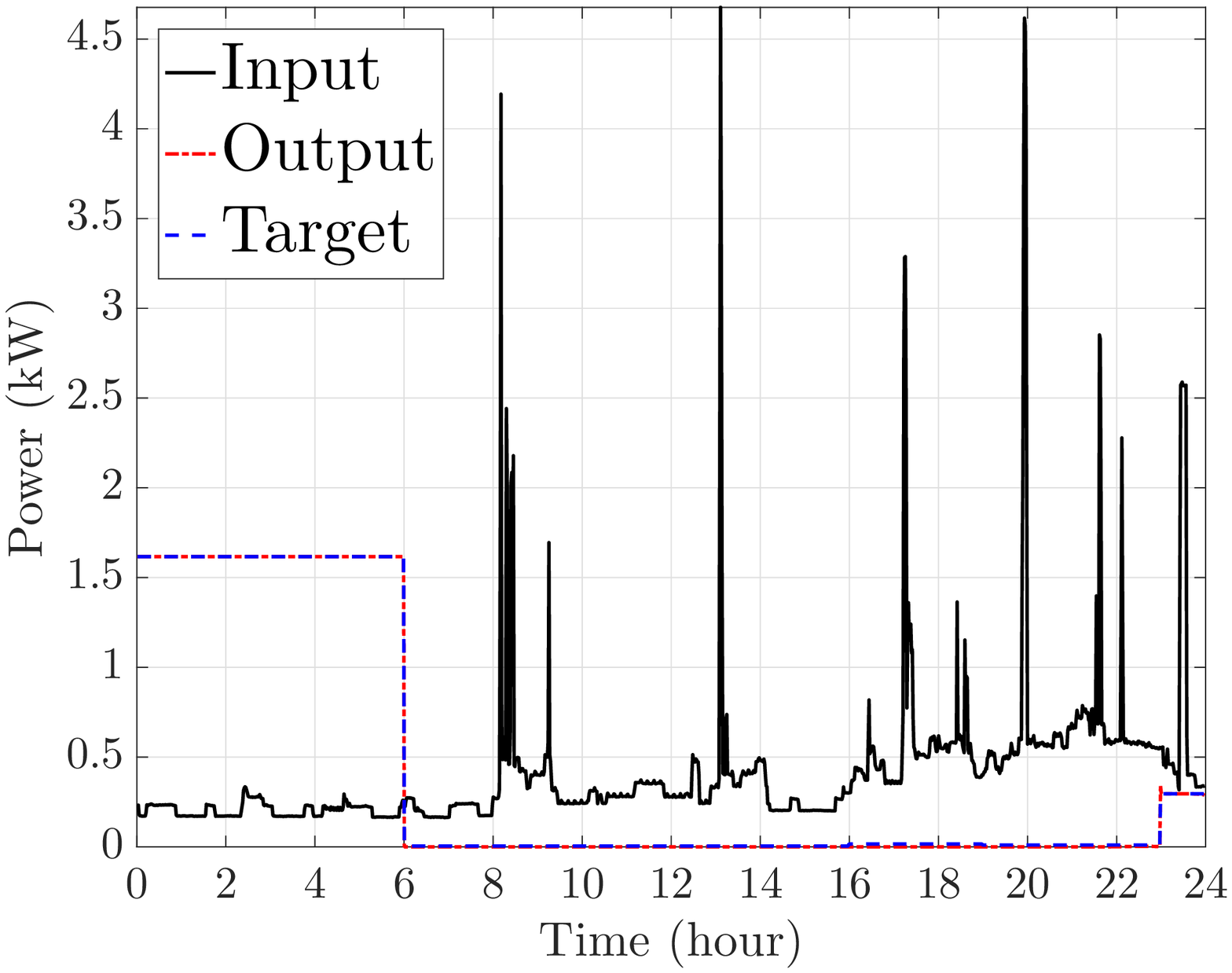}\label{fig:teslaNoSelling}}
\subfloat[][Energy can be sold.]
{\includegraphics[width=.5\columnwidth]{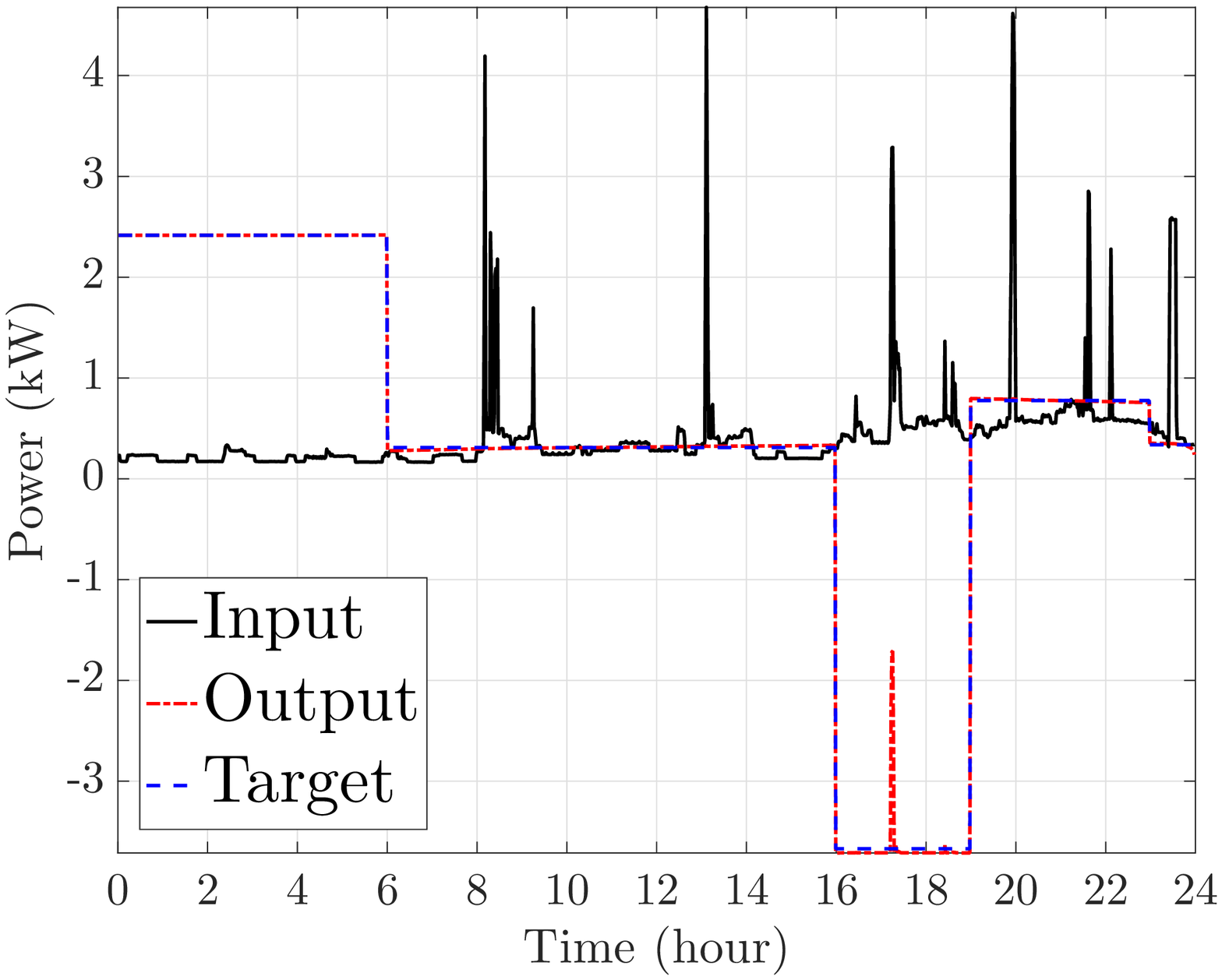}\label{fig:teslaSelling}}
\caption{Comparison between allowing to sell energy to the UP and not allowing it for the piecewise constant target load, a Tesla Powerwall 2 \cite{tesla} and $\alpha=0.5$. }
\label{fig:SellingVSnoselling}
\end{figure}

\begin{figure}[!t]
\centering
\includegraphics[width=.77\columnwidth]{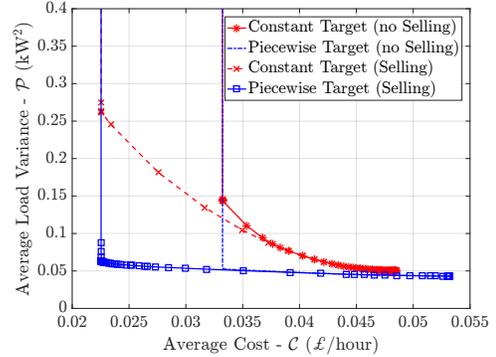}
\caption{$\mathcal{P}$-$\mathcal{C}$ trade-off when using a Powervault G200-LI-4KWH battery.}
\label{fig:tradeoff}
\end{figure}

In this section we evaluate numerically the performance achieved when using the piecewise constant target and compare it with the constant energy consumption target in \cite{Tan:2017}. We note that we also include the peak power constraints (\ref{eq:peakPowerCharging})-(\ref{eq:peakPowerDischarging}) and the total energy constraint (\ref{eq:totalEnergy}). We use real power consumption traces retrieved from the UK Dale dataset \cite{UK-DALE}, which has a time resolution of $6$ seconds. For our simulations, we convert the dataset readings to a time resolution of $1$ minute to reduce the computational complexity. We adopt two commercial batteries, listed in Table \ref{tab:batteryCapacity}, in which the charging (discharging) peak power denotes the amount of power that the battery is able to input (output) sustainably over time. We consider a time-of-use tariff currently being offered in the UK \cite{Tide:2017}, according to which the off-peak price is $4.99$p per kWh during 23:00 to 06:00, the medium price is $11.99$p per kWh during 06:00 to 16:00 and during 19:00 to 23:00, and the peak price is $24.99$p per kWh during 16:00 to 19:00.


Fig. \ref{fig:profiles} shows the input, output and target curves according to the two target loads for $\alpha=0.1$ in Fig. \ref{fig:profilesConstant01_B4} and Fig. \ref{fig:profilesPiece01_B4}, and $\alpha=0.9$ in Fig. \ref{fig:profilesConstant09_B4} and Fig. \ref{fig:profilesPiece09_B4}, respectively. These weights are chosen to simulate users that are mainly interested in minimizing their cost and maximizing privacy, respectively. For these figures we adopt a Powervault G200-LI-4KWH battery \cite{powervault}, and we assume that energy cannot be sold to the UP. As the figures show, adopting a piecewise constant target reduces the variance of the output load around the target, thanks to the flexibility introduced by the piecewise constant target function. However, because of the rather small battery capacity, the demand peaks in the input load are mostly revealed in the output load for both strategies. When $\alpha=0.1$, since cost is more important than privacy, the battery is charged more during the off-peak price period, and the outputs for the two scenarios are more similar, since reducing the variance from the target assumes less importance. 

Fig. \ref{fig:SellingVSnoselling} shows the output and target loads when selling energy to the UP is allowed, and a piecewise constant target is considered along with a Tesla Powerwall 2 \cite{tesla} and $\alpha=0.5$. The results show that the target in this case can also have negative values, which allow the user to sell energy to the grid during the peak price period. Selling energy to the grid further minimizes the costs and improves the overall trade-off.

Fig. \ref{fig:tradeoff} shows the trade-off between privacy and costs for the two strategies by considering the Powervault G200-LI-4KWH \cite{powervault}, and including the setting in which energy can be sold. The figure confirms the results of Fig. \ref{fig:SellingVSnoselling} by showing that, when energy can be sold, much lower costs are achieved for the same level of privacy. As expected, adopting a piecewise constant target leads to a much better privacy-cost trade-off compared to the constant target, both when selling and not selling energy.


\begin{figure}
\centering
\subfloat[][$\mathcal{P}$ versus $B_{\max}$ for $\alpha=1$.]
{\includegraphics[width=0.5\columnwidth]{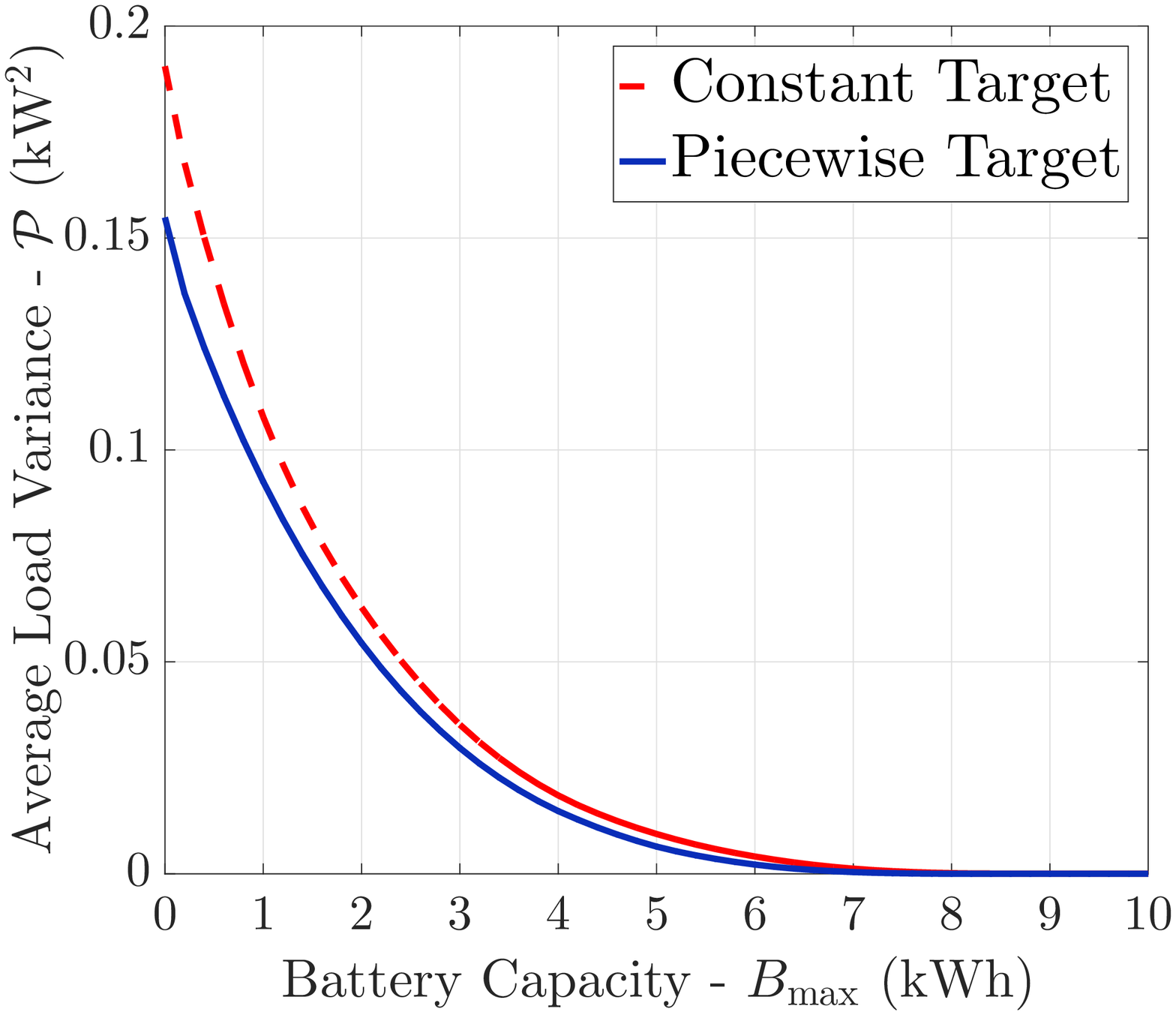}\label{fig:varianceVSbattery}}
\subfloat[][$\mathcal{C}$ versus $B_{\max}$ for $\alpha=0$.]
{\includegraphics[width=0.5\columnwidth]{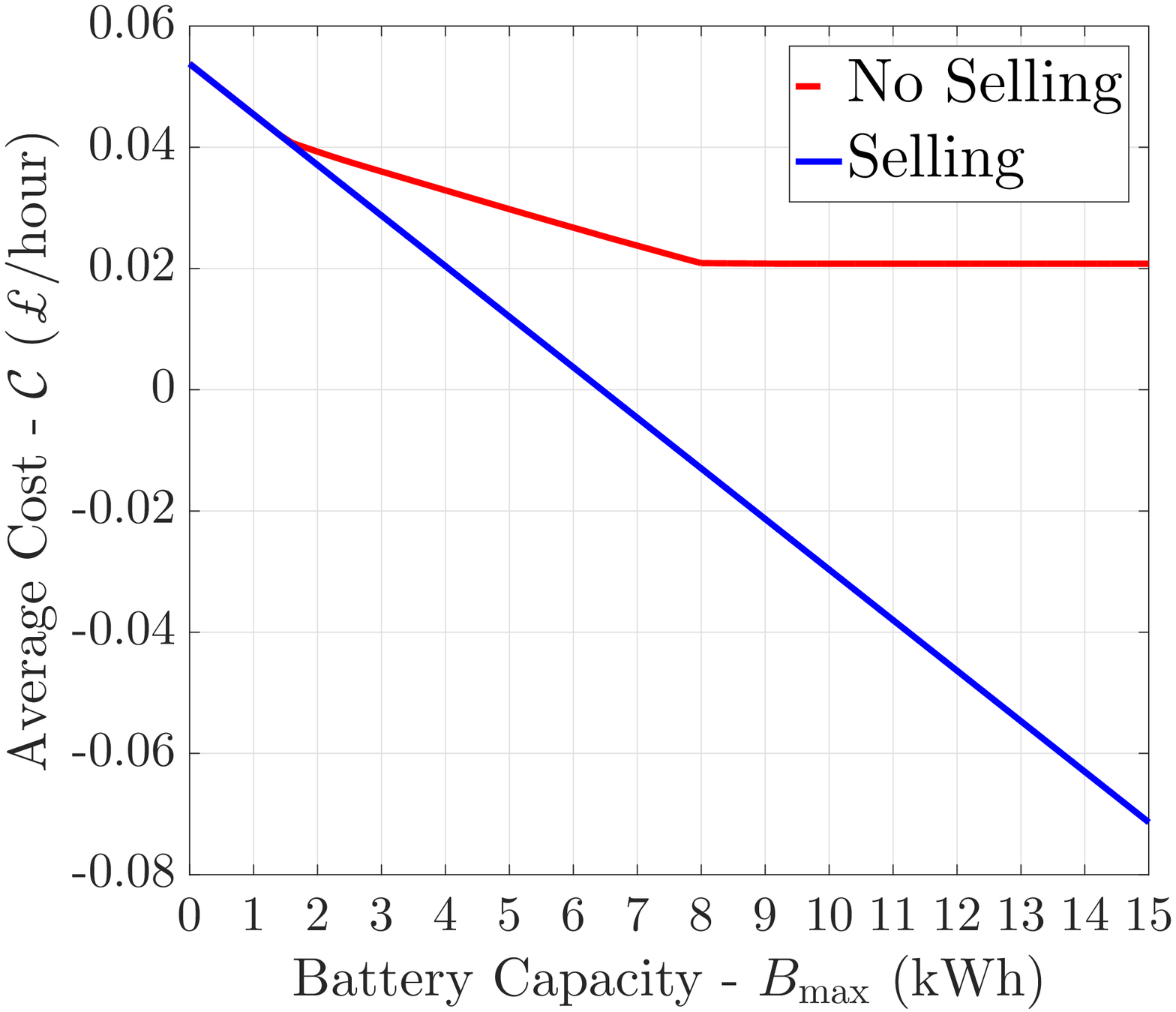}\label{fig:costVSbattery}}
\caption{Privacy and cost versus battery capacity.}
\label{fig:cosVarianceVSbattery}
\end{figure}



Fig. \ref{fig:varianceVSbattery} shows the degree of privacy that can be achieved for different battery capacities for $\alpha=1$, while Fig. \ref{fig:costVSbattery} shows the hourly average energy cost with respect to $B_{\max}$ for $\alpha=0$. For these plots, we set $\hat{P}_c=\hat{P}_d=B_{\max}/2$ which is a good approximation of the specifications of currently available residential batteries. Fig. \ref{fig:varianceVSbattery} shows that the information leakage drops greatly even for relatively small battery sizes, and it saturates at around $B_{\max}=8$ kWh for all the strategies considered. This shows that, if a user is only interested in her privacy and has a medium-sized battery ($\geq 6$ kWh) available, she can achieve almost zero output load variance by using any of the target loads considered. Also, for smaller battery capacities the piecewise constant target achieves better performance. However, we remark here that having multiple energy levels has an inherent information leakage which is not captured solely by the load variance from the target. It is noteworthy that selling energy does not improve privacy when $\alpha=1$, since selling introduces negative target values which are more difficult to match with the positive output load unless a large battery is used. In the setting of Fig. \ref{fig:costVSbattery}, as privacy is not relevant, the only difference between the two targets is due to allowing or not to sell energy to the UP. As one would expect, the flexibility to sell energy to the grid can significantly reduce the energy cost, and in principle the user can even make a profit (corresponding to a negative cost in the figure).


\section{Conclusions} \label{sec:conclusion}

In this paper we have studied the joint optimization of privacy and cost for an SM system with a rechargeable storage device. Privacy is measured via the mean squared-error between the SM measurements and a target load profile, which is set to be a piecewise constant function over price periods. The user is allowed to sell excess energy to the UP. The optimal solution of the corresponding optimization problem has been characterized, highlighting its water-filling interpretation. Detailed numerical simulations have been presented, showing the performance improvement of the proposed piecewise target load as compared to a constant target load. Future work will consider solving the corresponding online optimization problem, including the possibility of load shifting, while another interesting research direction is how to balance this framework with UP's interest in stabilizing the grid.

\bibliographystyle{IEEEtran}
\bibliography{SGC_reviewed}
\end{document}